# Global haze aerosol distribution: a direct view by Geofen-5 satellite with 3.3 km spatial resolution

Zhengqiang Li[1], Yisong Xie[1,*], Weizhen Hou[1], Hua Xu[1], Kaitao Li[1], Li Li[1], Yang Zhang[1,2]

[1] State Environment Protection Key Laboratory of Satellite Remote Sensing, Institute of Remote Sensing and Digital Earth, Chinese Academy of Sciences, Beijing 100101, China

[2] College of Resources and Environment, Chengdu University of Information Technology, Chengdu 610225, China

*Corresponding author, E-mail: xieys@radi.ac.cn

**Abstract:** The Directional Polarimetric Camera (DPC) is the first Chinese multi-angle polarized earth observation satellite sensor, which has been launched onboard the GaoFen-5 Satellite in Chinese High-resolution Earth Observation Program. GaoFen-5 runs in a sun-synchronous orbit with the 2-days revisiting period. DPC employed a charge coupled device detection unit, and can realize spatial resolution of 3.3 km under a swath width of 1850 km. Moreover, DPC has 3 polarized channels together with 5 non-polarized bands, and is able to obtain at least 9 viewing angles by continuously capturing series images over the same target on orbit. Based on the Directional Polarization Camera (DPC) onboard GF-5 satellite, the first global high-resolution (3.3 km) map of fine-mode aerosol optical depth (AODf) over land has been obtained together by Aerospace Information Institute, Chinese Academy of Sciences, Anhui Institute of Optics and Fine Mechanics, Chinese Academy of Sciences (the manufacturer of DPC sensor) and other institutes. This AODf remote sensing observation dataset has the highest spatial resolution in the world. It can reflect the spatial information of major air pollutants (PM2.5, etc.) and provide critical basic products for "decryption" of global haze distribution.

# 1. Introduction

Aerosol is an important component of the global atmosphere and has a significant impact on climate change, air pollution, material transport, and ecological assessment. The aerosol components, such as black carbon deposited on ice and snow, further accelerate the ablation of ice and snow, and generate feedback-coupling effect between the cryosphere, hydrosphere, and radiation circle in the polar region (Xie et al., 2018). These complex effects make aerosols, including their cloud effects, become the most important factor affecting the uncertainties in the climate change assessment (Mishchenko et al., 2004). On the other hand, aerosols with aerodynamic diameters of less than 2.5 microns (PM2.5) have become the focus of attention for atmospheric pollution, because they can enter the people's lung with breathing and affect vital systems such as blood circulation (Zhang and Li, 2015; Li et al., 2016). Thus, it is very important to obtain spatial distribution of aerosol parameters in large-scale and even global scales based on the satellite remote sensing (Mishchenko et al., 2007).

Many satellite platforms enabled the retrieval of the first optical parameter of aerosols -- aerosol optical depth (AOD) from the top of atmosphere (TOA) reflectance. Among those achievements, the Moderate-resolution Imaging Spectroradiometer (MODIS) AOD product over land, which is retrieved by the dark target (DT) algorithm, has a good retrieval accuracy and is widely used in the meteorological and environmental area (Levy et al., 2007, 2013). One of the most famous application of AOD is the modeling of the particulate matter (PM) concentration, and recent studies developed a physical PM2.5 remote sensing (PMRS) model that has the features of fast computation and easy implementation (Zhang and Li, 2015; Li et al., 2016). An important input parameter in the PMRS model is the fine-mode AOD (AODf), which is obtained by using the aerosol fine-mode fraction (FMF) in the current approach. Although the MODIS platform provides the FMF product, its retrieval accuracy over land is poor (Levy et al., 2010), and it consequentially limits the PM2.5 estimation precision obtained from the PMRS model. Specifically, if we can retrieve AODf from satellite observation directly and precisely instead of using FMF, the model accuracy should be improved. Moreover, AODf is also important in the field of global climate change, because it basically represents the anthropogenic aerosols component, which can be used to compute the radiative forcing of anthropogenic aerosols (IPCC, 2014). Therefore, AODf is a meaningful aerosol optical parameter that deserved to be retrieved from the satellite observation.

The Polarization and Directionality of Earth's Reflectance (POLDER) and Polarization and Anisotropy of Reflectances for Atmospheric Science coupled with Observations from a Lidar

(PARASOL) instruments, which have an ability to detect the polarized light in addition to the traditional intensity measurement, provide an opportunity to retrieve more aerosol optical and physical parameters (Deuzé et al., 2001; Tanré et al., 2011). Because the polarization signal received by the POLDER/PARASOL sensor mainly comes from the radiation contribution of fine-mode aerosols, and the coarse-mode aerosols give a negligible contribution, this feature is directly used in the POLDER/PARASOL operational algorithm for AODf retrieval over land (Deuzé et al., 2001; Tanré et al., 2011; Herman et al., 1997). In particular, the global long-term sequences of multi-angle polarization datasets were obtained by the PARASOL from the year of 2004 to 2013. By developing the corresponding algorithms, many research breakthroughs have been made, such as the multi-parameter inversion of aerosols on the ocean (Hasekamp et al., 2011), the inversion of fine aerosols over land, and the observations of the interaction between clouds and aerosols (Costantino et al., 2010). In recent years, with the progress of statistical optimization methods (Dubovik et al., 2011) and polarized surface models (Nadal et al., 1999; Maignan et al., 2009; Waquet et al., 2009), a new generation of multi-parameter POLDER inversion method has basically matured and applied to the reprocessing of POLDER historical data.

Based on the POLDER technology, the Anhui Institute of Optics and Fine Mechanics belonging to Chinese Academy of Sciences, has developed a Directional Polarimetric Camera (DPC) space-borne sensor supported by the National Space Administration of China, with highlights of one time increase in spatial resolution from about 6 km × 7 km to 3.3 km × 3.3 km (Li et al., 2018). The first DPC sensor has been launched at May of 2018 with the Chinese atmospheric environment flagship satellite GaoFen-5 (GF-5) in the Chinese High-resolution Earth Observation System (CHEOS) program (Gu et al., 2015), and a series of data has been obtained overworld. Thus, it is necessary to retrieve the global high-resolution map of fine-mode aerosol optical depth (AODf) over land from DPC/GF-5, and further to monitor the global haze distribution by satellite remote sensing. In response to this requirement, this paper firstly introduced the GF-5 satellite and DPC sensor in Section 2, then established the methodology including retrieval algorithm and data processing method in Section 3. After that, the retrieved global AODf results and validations are presented in Section 4. Finally, we get the conclusion of our work in section 5.

## 2. GF-5 satellite and DPC sensor

### 2.1 GF-5 satellite

As the flagship of the environment and atmosphere observation satellite in the Chinese High-resolution Earth Observation System (CHEOS) program (Gu et al., 2015), GF-5 has been launched in May 5, 2018. There are six payloads onboard the GF-5 satellite, in which one of important senor is Directional Polarization Camera (DPC), just as illustrated in Figure 1 (Li et al., 2018). The GF-5 satellite runs in a sun-synchronous orbit with an inclination angle of 98º. The overpass local time is 13:30 PM and the design life is 8 years. The six sensors onboard GF-5 works together for environment monitoring and assessment, focusing on aerosols, gas-house gases, polluted gases, cloud, water vapor, waste water, land surface, vegetation, biomass burning and urban environment, etc.

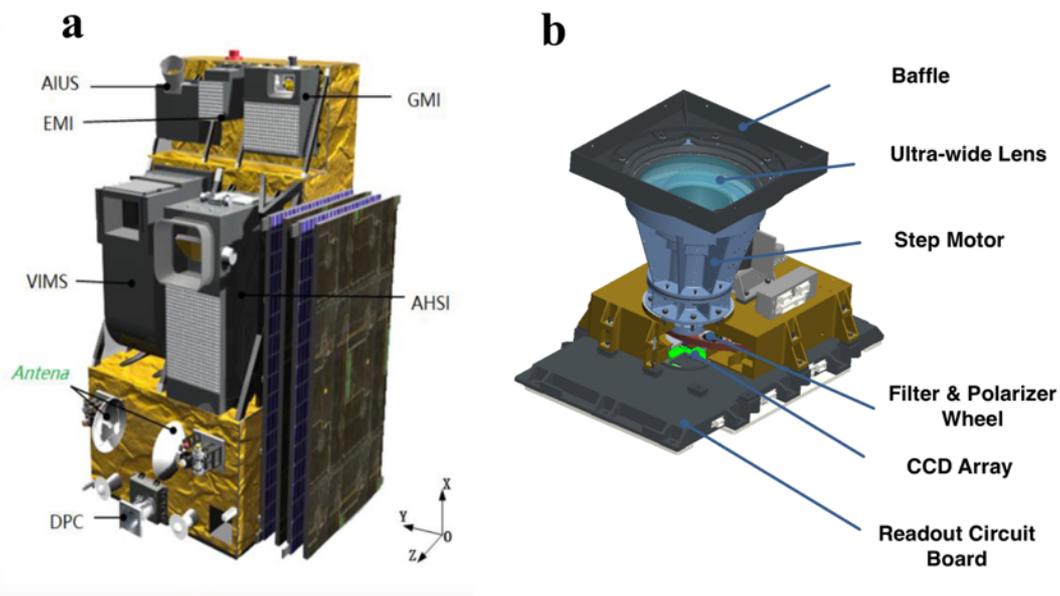

**Figure 1.** Illustration of the GF-5 satellite (a) and the Directional Polarized Camera (b).

### 2.2 DPC sensor

The Directional Polarimetric Camera (DPC) is the first Chinese multi-angle polarized earth observation satellite sensor, which is the type of POLDER polarimetric imager. The DPC employed a charge coupled device (CCD) detection unit with 512×512 effective pixels from the 544×512 useful pixels, and can realize spatial resolution of 3.3 km under a swath width of 1850 km. The 2-days revisiting period allows it visiting the temporal variation of atmospheric pollution. Table 1 lists the instrumental parameters of DPC sensor on board GF-5 detailedly.

There are 3 polarized bands (490, 670 and 865nm) and 5 non-polarized bands (443, 565, 763, 765 and 910 nm) integrated in DPC sensor, and the spectral widths of these 8 bands are from 10 to 40 nm. The radiometric and polarimetric calibration errors are less than about 5% and 0.02 (with respect to the degree of linear polarization, abbreviated as DOLP which is in a range of 0−1) respectively, which will be continuously maintained by performing on-orbit vicarious calibration combing laboratory calibration coefficients. DPC share many similarities in the instrument design with POLDER, the distinguishing feature is that the pixel spatial resolution of DPC (3.3 km × 3.3 km) has significant improvement than POLDER (about 6 km × 7 km), thus higher spatial resolution of aerosol and surface products could be further retrieved from the measurements of DPC (Li et al., 2018).

Table 1. The instrumental parameters of DPC sensor on board GF-5 satellite

| Parameter | Value | Parameter | Value |
| --- | --- | --- | --- |
| Instrument FOV | ±50° (across/along-track) | Polarized angle | 0°, 60°, 120° |
| Spatial res. (km) | 3.3 | Stokes parameters | I, Q, U |
| Swath width (km) | 1850 | Rad. Cal. Error | ≤ 5% |
| Multi-angle | ≥ 9 | Pol. Cal. Error | ≤ 0.02 |
| Image pixels | 512×512 | Band width (nm) | 20, 20, 20, 20, 10, 40, 40, 20 |
| Spectral band (nm): P for polarization | 443, 490(P), 565, 670(P), 763, 765, 865(P), 910 | | |

## 3. Methodology

### 3.1 Retrieval algorithm

The retrieval theory of fine-mode aerosol optical depth (AODf) had been detailedly introduced in the study of Zhang et al. (2016, 2017, 2018), we directly discuss the existing shortcomings here.

In the multi-angular polarized aerosol retrieval approach, a series of aerosol models are used to simulate the top of atmosphere (TOA) polarized reflectance, and a merit function are applied to determine the optimal aerosol model that best fits the observation. Most merit functions are based on calculating the accumulated residual error between the simulated and observed multi-angle TOA radiation. For example, the method of determining the optimal aerosol model in the DPC operational algorithm can be expressed as

$$\eta = \sqrt{\frac{1}{2N}\Sigma_{\lambda_0,\lambda_1}\Sigma_j[R_{cal}(\lambda,\Theta_j) - R_{meas}(\lambda,\Theta_j)]^2}, \quad (1)$$

where $\eta$ is the accumulated residual error; $N$ is the number of observation angles; $\lambda_0$ and $\lambda_1$ are the PARAOSL 670 nm and 865 nm bands, respectively; $\Theta$ is the scattering angle; $R_{cal}(\lambda,\Theta_j)$ means the calculated polarized reflectance for each $\lambda$ and $\Theta$; $R_{meas}(\lambda,\Theta_j)$ means the observed polarized reflectance of the corresponding $\lambda$ and $\Theta$. Because each aerosol model has an optimal AODf by comparing the simulated and observed TOA polarized reflectance for each observation angle, and the smallest $\eta$ can distinguish the optimal aerosol model, then the AODf and aerosol model can be determined.

Although this merit function has an obvious physical meaning, the observation error is not considered. It should be noted that Eq. (1) shows a process that accumulating the residual error for multi-angle, the more angles, the more observation errors were disregarded, which would eventually lead to an inappropriate selection of aerosol models. In other words, the accumulated observation errors may amplify or mask the retrieval errors, but it is hard to know the concrete effect on the retrieval errors, which lead to a fact that the model with the smallest residual error is not necessarily the optimal model. To solve this problem, we further introduce the grouped residual error sorting (GRES) method for optimal aerosol model determination (Zhang et al., 2018).

### 3.2 Date processing

Generally, the polarized signals are mainly produced by the fine-mode aerosols in the range 80°<$\Theta$<120°, so we only employ the data with $\Theta$ in that range to retrieve AODf from the polarization measurements of DPC. The forward model for aerosol retrieval can be expressed as:

$$R_{pol}^{TOA} = R_{pol}^{atm} + R_{pol}^{surf} \cdot exp(-M\tau_m - Mc\tau_a), \quad (2)$$

where $R_{pol}^{TOA}$ is the TOA polarized reflectance; $R_{pol}^{atm}$ is the atmospheric polarized contribution, which is mainly composed of the polarized aerosol and molecule reflectance, and the polarized aerosol reflectance is mainly generated by the fine-mode spherical particles, the coarse-mode contribution is neglected; $R_{pol}^{surf}$ is the surface polarized reflectance; $M$ is the air mass; $\tau_m$ is the molecular optical depth; $\tau_a$ is the fine-mode AOD, and $c$ accounts for the large forward scattering of the aerosol (Deuzé et al., 2001; Zhang et al., 2018).

The concrete steps of the data processing are as follows:

Step 1: Set a 3×3 window, begin the retrieval process if all pixels in this window are clear, otherwise not.

Step 2: Obtain the observation geometry parameters of the pixel from the satellite data, and then search the values that are close or equal to the observations in the LUT.

Step 3: Input all the atmospheric polarized reflectance values that match the observation geometry parameters in the LUT into Eq. (2) to obtain the two closet sets of atmospheric polarized parameters and the corresponding AODs.

Step 4: Perform a linear interpolation for the atmospheric polarized parameters according to the observation geometry parameters and AODs, and then a new high dimension LUT is generated.

Step 5: Input the results of the new LUT into Eq. (2) to execute a comparison with the observed TOA polarized reflectance; derive the AODf whose simulated TOA polarized reflectance is closest to the observation as the optimal AODf of the corresponding aerosol model.

Step 6: Repeat the four steps above, and then the 25 sets of AODf and residual errors can be obtained.

Step 7: Use the GRES method to obtain the final AODf.

## 4. Results and validation

### 4.1 Retrieval results and comparisons

Figure 2 shows the global distribution of mean AODf results at 865nm over land retrieved by DPC/GF-5 with about 3.3km resolution during November 23 to November 30 in 2018, in which red denotes high AODf regions, and light grey indicates cloud regions or lack of valid data. It should be noted that the retrieved AODf results from DPC/GF-5 are the world's highest spatial resolution products by the means of satellite remote sensing, which is about 6 times higher than the official product of POLDER/PARASOL.

Figure 3 illustrates the AODf distribution and pixel resolution retrieved by DPC/GF-5 and the official products by POLDER/PARASOL in Beijing, respectively. For the reason that the POLDER/PARASOL has been no longer working in orbit after 2013, so we show the retrieved results in the same month but in different year. In Figure 3, left panel of shows the retrieved AODf with 3.3km spatial resolution by DPC/GF-5 in November 2018, while right panel shows the official AODf with 18.5km spatial resolution by POLDER/PARASOL in November 2011. The high spatial resolution AODf products can clearly show the local details of pollution distribution, and further support for the regional pollution fine control, key city pollution transmission monitoring, pollutant traceability and other environmental protection services.

Figure 4 further show the retrieved AODf results by DPC/GF-5 and the official products by POLDER/PARASOL in some key global regions, in which the East China, India and central Africa are considered respectively. In which, the left panel represents the retrieved AODf from DPC/GF-5 measurements with 3.3km spatial resolution in November 2018, while the right panel represents the official AODf products from POLDER/PARASOL measurements with 18.5km spatial resolution in November 2011.

From the comparison of global key regions in the longer time span (2018 versus 2011), it can be seen that the pollution level in eastern China has been significantly decreased compared with the peak situation in November 2011, especially in the southeast coastal region, there is a significant reduction. However, the fine-mode aerosols in northern China is still high and need further control and improvement. In contrast, India, another key polluted area in the world, has shown significant pollution growth, reflecting the increase in human activities such as increased industrial and agricultural emissions. In addition, the pollution situation in central Africa is also significantly enhanced compare to the year of 2011, unlike these regions with rapid economic growth, the changes in AODf in this region are mainly affected by factors such as natural biomass burning and forest destruction.

### 4.2 Preliminary validation

In response to major requirement such as PM2.5 remote sensing estimation, the applicability function of fine-mode truncation module is added to the inversion algorithm in order to further improve the correlation between satellite products and PM2.5, and some preliminary validation have been carried out.

Figure 5 show the validation results of retrieved AODf from DPC/GF-5, in which panel (a) illustrates the global Aerosol Robotic Network (AERONET) (Holben et al., 1998) and Sun-Sky Radiometer Observation Network (SONET) (Li et al., 2018) sites for the validations of AODf satellite products; panel (b) plots the used ground PM2.5 observation sites in Beijing; panel (c) shows the correlation between retrieved AODf and ground monitoring PM2.5.

The synchronized validation data for AODf retrievals from 215 ground-based stations worldwide show that the satellite remote sensing results of DPC have great consistency with ground-based observations. Meanwhile, the joint analysis with surface PM2.5 data in Beijing shows that satellite remote sensing AODf has a good correlation with PM2.5 concentration, indicating the good potential of DPC in quantitative estimation of PM2.5.

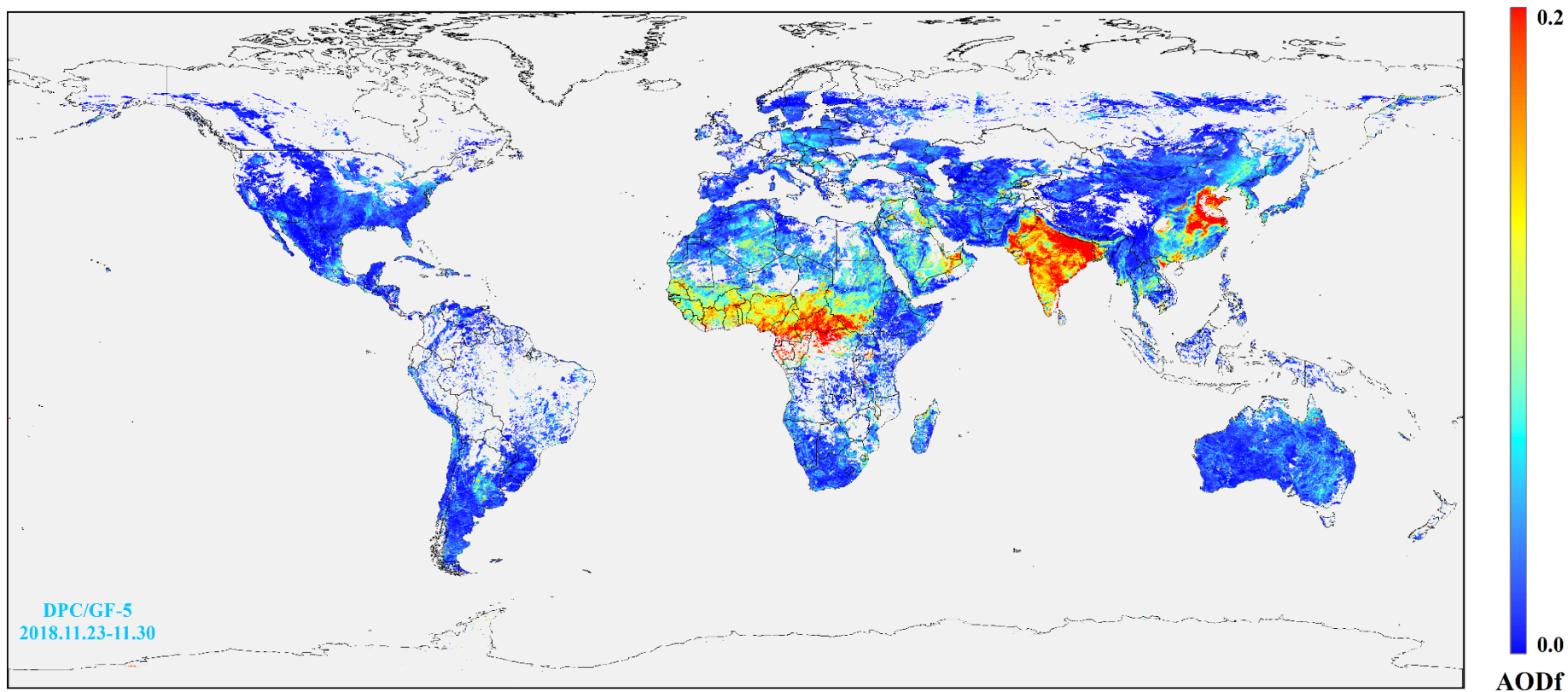

**Figure 2.** Global distribution of fine-mode aerosol optical depth (AODf) over land retrieved by DPC/GF-5. Red denotes high AODf regions, and light grey indicates cloud regions or lack of valid data.

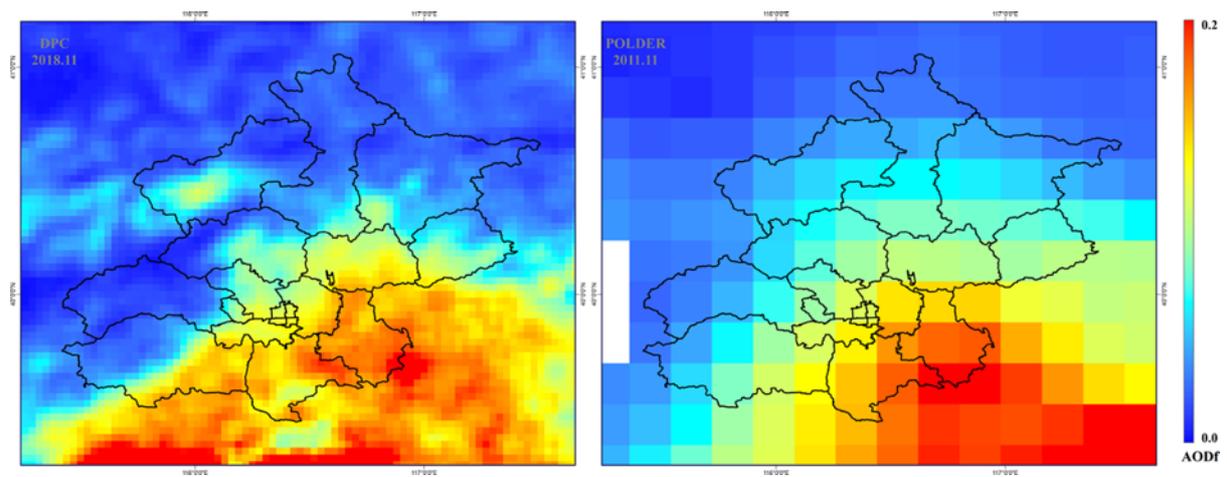

**Figure 3.** The retrieved AODf distribution and pixel resolution by DPC/GF-5 and the official AODf products by POLDER/PARASOL in Beijing. Left panel: the retrieved AODf with 3.3km spatial resolution by DPC/GF-5 in November 2018; right panel: the official AODf products with 18.5km spatial resolution by POLDER/PARASOL in November 2011.

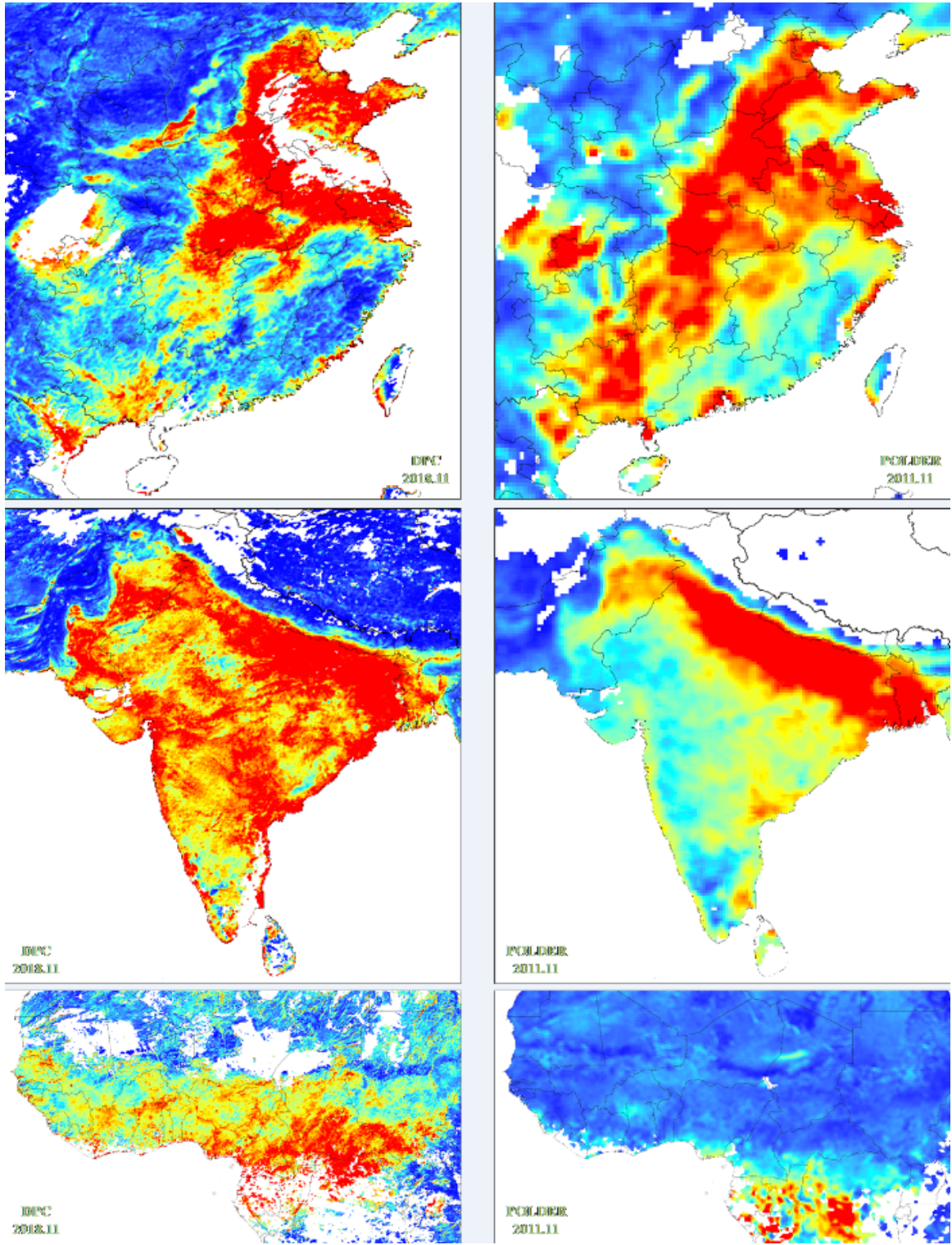

**Figure 4.** The retrieved AODf results by DPC/GF-5 and the official products by POLDER/PARASOL in some key global regions, in which the top, middle and lower corresponds to East China, India and central Africa, respectively. The left panel represents the retrieved AODf from DPC/GF-5 measurements with 3.3km spatial resolution in November 2018, while the right panel represents the retrieved official AODf products from POLDER/PARASOL measurements with 18.5km spatial resolution in November 2011.

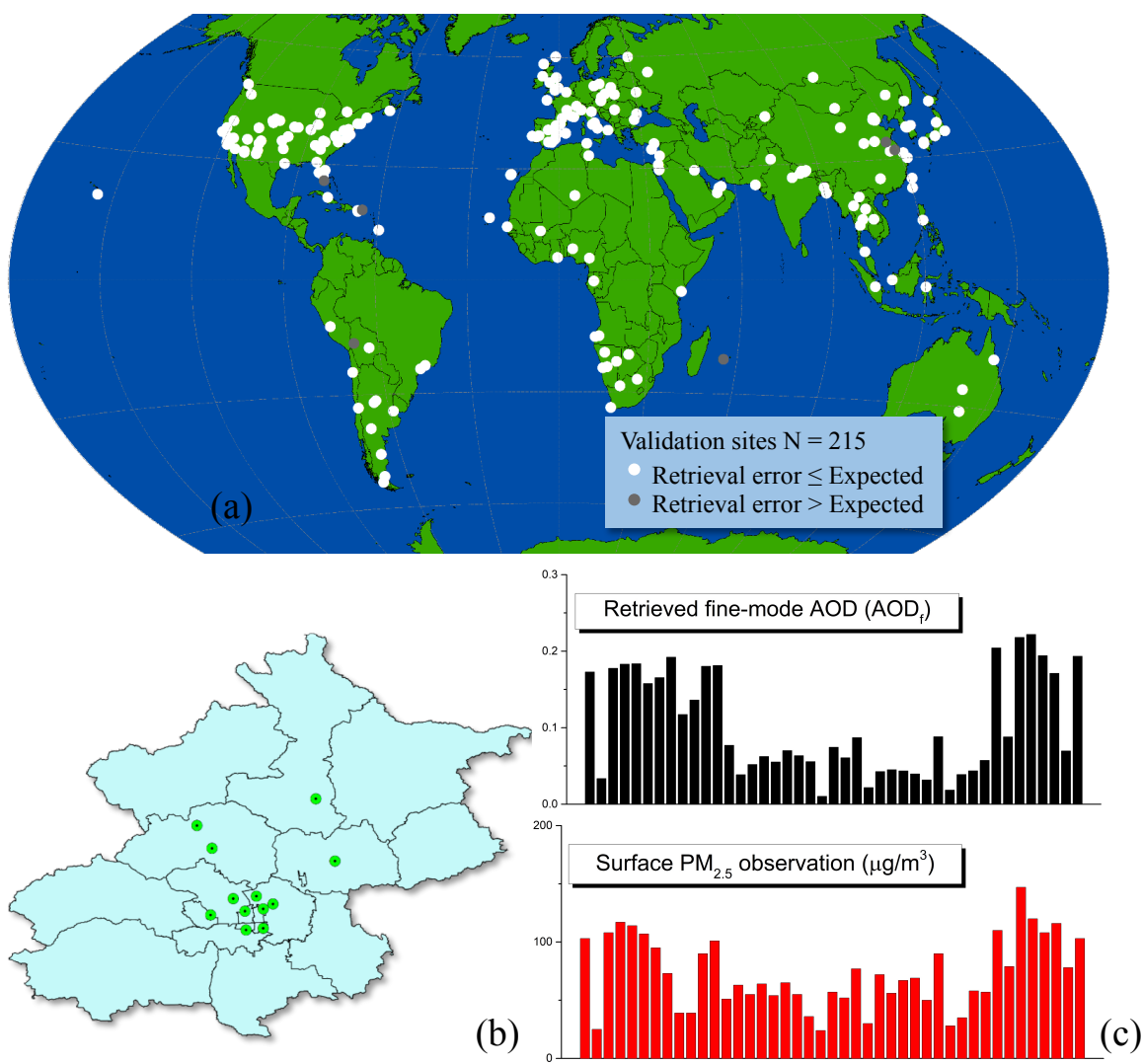

**Figure 5.** (a) Global AERONET and SONET sites for the validations of AODf satellite products; (b) The used ground PM2.5 observation sites used in Beijing; (c) There is a good correlation between retrieved AODf and ground monitoring PM2.5.

## 5. Conclusions

Based on the Directional Polarization Camera (DPC) onboard GF-5 satellite, the first global high-resolution (3.3 km) map of fine-mode aerosol optical depth (AODf) over land has been obtained. This AODf remote sensing observation dataset has the highest spatial resolution in the world. It can reflect the spatial information of major air pollutants (PM2.5, etc.) and provide critical basic products for "decryption" of global haze distribution.

Moreover, the AODf retrieved by satellite remote sensing is the total optical quantity of fine particles of the entire atmosphere, while PM2.5, which is closely related to environment and human health, refers specifically to the mass concentration of fine particles in the surface layer. Therefore, accurate estimation of surface PM2.5 by satellite AODf still requires a lot of research and innovation work (Zhang and Li, 2015; Li et al., 2016).